\documentclass[12pt]{article}
\renewcommand{\baselinestretch}{1.4}
\begin{document}
\begin{center}
{\Large \bf Diffusive Growth of a Single Droplet with Three Different Boundary Conditions}
\vskip 0.3cm

{\large Z.~Tavassoli \footnote{Electronic mail address: zohreh.tavassoli@brunel.ac.uk} and G.~J.~Rodgers}

\vskip 0.2cm
{\it Department of Mathematical Sciences, Brunel University,\\
Uxbridge, Middlesex, UB8 3PH, U.K.}
\end{center}
\vskip 0.0cm

{\noindent \large Abstract}

We study a single, motionless three-dimensional droplet growing by adsorption of diffusing monomers on a 2D substrate. The diffusing monomers are adsorbed at the aggregate perimeter of the droplet with different boundary conditions. Models with both an adsorption boundary condition and a radiation boundary condition, as well as a phenomenological model, are considered and solved in a quasistatic approximation. The latter two models allow particle detachment.
In the short time limit, the droplet radius grows as a power of the time with exponents of $1/4$, $1/2$ and $3/4$ for the models with adsorption, radiation and phenomenological boundary conditions, respectively.
In the long time limit a universal growth rate as $[t/\ln(t)]^{1/3}$ is observed for the radius of the droplet for all models independent of the boundary conditions. This asymptotic behaviour was obtained by Krapivsky \cite{krapquasi} where a similarity variable approach was used to treat the growth of a droplet with an adsorption boundary condition based on a quasistatic approximation.   
Another boundary condition with a constant flux of monomers at the aggregate perimeter is also examined. The results exhibit a power law growth rate with an exponent of $1/3$ for all times. 



\newpage
\renewcommand{\baselinestretch}{2}

\section{Introduction}
\label{sec:introduction}

There has been considerable interest in diffusive-growth processes including growth phenomena for a droplet on a substrate. This growth phenomena in the case where diffusion and coalescence play the major roles are common in many areas of science and technology \cite{sc1,sc2}. In this process, each droplet diffuses and grows individually and coalesces with contacting droplets. The kinetics of these phenomenon have been studied experimentally and theoretically [3-29]. Some models have been developed to explain the kinetics of these processes. One such model \cite{static88,static91,krapquasi} consists of a single, motionless three dimensional droplet formed by diffusion and adsorption of non-coalescing monomers on a 2D substrate. In the model it is assumed that the diffusing monomers coalesce only with large immobile growing droplet and not with each other. In \cite{static88,static91} a {\it static} approximation was used to solve the diffusion equation and an approximate description of the long time behaviour was obtained. The static approach predicted an asymptotic power law growth rate for the radius of the droplet. Because of the growth of the droplet, the present problem involves  a moving boundary. Moving boundary problems in the context of the diffusion equation are referred to as Stefan problems [30-32]. The only exact solutions for these problems have been found using a {\it similarity variable} method, see for instance, [30-34] and references therein. Using this method for a droplet of dimensionality $d$ growing on a substrate of the same dimensionality, an exact scaling solution can be found. In \cite{harriso} such a solution in one dimension has been derived which can be generalised to a higher dimension. However, the problem of a 3D droplet growing on a 2D substrate, may be treated by approximate methods. A simple treatment based on a {\it quasistatic} approximation has been presented in \cite{krapquasi}. A {\it similarity variable} approach was used to solve the Stefan problem with moving boundary \cite{krapquasi}. The results predicted that the radius of the droplet increases as $[t/\ln(t)]^{1/3}$ asymptotically. The asymptotic growth law predicted by the static approach differs from the quasistatic answer by a slowly varying logarithmic factor. In all the models in \cite{static88,static91,krapquasi} an adsorption boundary condition at the aggregate perimeter of the droplet, was considered. 

In \cite{mozy} a generalisation of Smoluchowski model \cite{smol,smolhelp} for diffusional growth of colloids, was presented. Smoluchowski \cite{smol} considered the process of diffusional capture of particles assuming the growing aggregate is modeled as a sphere. He then solved the diffusion equation with an absorbing boundary condition at the aggregate surface of the sphere. In \cite{mozy} two other approaches were considered, a phenomenological model for the boundary condition and a radiation boundary condition. Both approaches allowed for incorporation of particle detachment in Smoluchowski model. Explicit expressions for the concentration and intake rate of particles were given in the long time limit \cite{mozy}.

In this paper we consider a single, motionless three-dimensional droplet growing by adsorption of diffusing monomers on a two-dimensional substrate. The diffusing non-coalescing monomers are adsorbed at the aggregate perimeter of the droplet with different boundary conditions. Models with different boundary conditions for the concentration of monomers are considered and solved in a quasistatic approximation. For each model, the diffusion equation is solved exactly, subject to a {\it fixed} boundary. Using mass conservation law at the aggregate perimeter of the growing droplet, we then obtain an expression for the growth rate of the {\it moving} boundary. Explicit asymptotic solutions in the both short and long time limits are given for the concentration, total flux of monomers at the perimeter of the growing droplet and for the growth rate of the droplet radius. This paper is organised as follows. In section 2, a model with an adsorption boundary condition is examined. In sections 3 and 4 we consider the two approaches which were introduced in \cite{mozy} to allow for particle detachment. A phenomenological model and a model with a radiation boundary condition are considered in sections 3 and 4, respectively. Another boundary condition which assumes a constant flux of monomers at the aggregate perimeter of the droplet, is also introduced in section 5. Finally, in section 6 we compare the results of different approaches and summarise our conclusions.           

\section{Growth Equations with Adsorption Boundary Condition}
\label{sec:Adsorption}

Consider an immobile three-dimensional droplet which is initially surrounded by monodisperse droplets. The droplet lies on a two-dimensional plane substrate on which the monomers diffuse. 
Monomers have the volume $V$ and diffuse with the diffusion constant $D$. Then, the concentration of monomers at point $r$ and at time $t$, $c(r,t)$, is described by the diffusion equation

\begin{equation}
\label{difequ}
\frac{\partial c(r,t)}{\partial t} = D\,\frac{1}{r}\,\frac{\partial}{\partial r}\left(r\,\frac{\partial c(r,t)}{\partial r}\right)
\end{equation}
for $r\geq R$, where $R(t)$ is the radius of the immobile growing droplet. The initial conditions are given by

\begin{equation}
\label{cinitial}
c(r,t=0) = c_0,
\end{equation}
which is the initial, uniform, monomer concentration and

\begin{equation}
\label{Rinitial}
R(t=0) = 0,
\end{equation}
which shows that the droplet is not present at the beginning of the process. We consider an adsorption boundary condition at the perimeter of the droplet

\begin{equation}
\label{adbc}
c(r=R,t>0) = 0
\end{equation}
and assume that at infinity the concentration of the monomers is finite and equal to $c_0$.
Concentration gradients in the neighborhood of the droplet create a flux of monomers on the two-dimensional substrate. This flux feeds the growth of the droplet. Therefore, the rate of increase of the droplet volume is related to the total flux of monomers at the perimeter of the droplet by mass conservation,

\begin{equation}
\label{growthdef}
\Phi (t) = \lambda \,R^2\,\frac{dR}{dt},
\end{equation}
where the total flux

\begin{equation}\label{intakedef}
\Phi (t) = V\left[2\pi RD\left. \frac{\partial c}{\partial r} \right\vert_R\right]
\end{equation}
corresponds to the monomers incorporated at the perimeter of the droplet.
In (\ref{growthdef}) $\lambda $ is a dimensionless factor related to the contact angle of the droplet.  

In order to solve (\ref{difequ}) with (2-4), we introduce the Laplace transform of the concentration,

\begin{equation}
\bar c(r,s) = \int_0^\infty dt\,e^{-st}c(r,t),
\end{equation}
which satisfies the equation

\begin{equation}
D\,\frac{1}{r}\,\frac{\partial}{\partial r} \left( r\,\frac{\partial \bar c}{\partial r}\right) = s\,\bar c - c_0.
\end{equation}
Here we have already used the initial condition (\ref{cinitial}). The general solution of this equation is given by

\begin{equation}
\bar c(r,s) = \frac{c_0}{s} + A(s)K_0(qr) + B(s)I_0(qr),
\end{equation}
where $q = \sqrt {s/D}$, and $K_0$ and $I_0$ are Modified Bessel functions of order zero. To have a finite solution as $r\to \infty $, we set $B(s) = 0$. The boundary condition (\ref{adbc}) in the Laplace transform version becomes

\begin{equation}
\label{cbar at r=R}
\bar c(r=R,s)=0.
\end{equation}
Using (\ref{cbar at r=R}) the transformed concentration and its gradient normal to the droplet perimeter, yield

\begin{equation}
\label{cbarad}
\bar c(r,s) = \frac{c_0}{s} \left[ 1 - \frac{K_0(qr)}{K_0(qR)} \right],
\end{equation}

\begin{equation}
\label{dcbarad}
\frac{\partial \bar c(r,s)}{\partial r} = \frac{c_0}{(Ds)^{1/2}} \,\frac{K_1(qr)}{K_0(qR)}.
\end{equation}
To find time dependent concentration and its radial gradient, we use the Inversion theorem for (\ref{cbarad},\ref{dcbarad}). Both (\ref{cbarad},\ref{dcbarad}) have a branch point at $s=0$, so in the Inversion formula, we use a contour which does not contain any zeros of $s$ and $K_0(qR)$. Consequently, time dependent concentration and also total flux at the droplet perimeter from (\ref{intakedef}), are given by

\begin{equation}
\label{cad}
c(r,t) = \frac{2c_0}{\pi}\int_0^\infty e^{-Du^2t} \left[ \frac{J_0(Ru)N_0(ru) - J_0(ru)N_0(Ru)}{J_0^2(Ru) + N_0^2(Ru)}\right] \frac{du}{u},
\end{equation}

\begin{equation}
\label{intakead}
\Phi (t) = \frac{8\,c_0DV}{\pi}\int_0^\infty e^{-Du^2t}\frac{1}{\left[ J_0^2(Ru) + N_0^2(Ru) \right]}\frac{du}{u},
\end{equation}
where $J_0$ and $N_0$ are Bessel functions of order zero.
Using (\ref{growthdef},\ref{intakead}) a differential equation for the growth rate of the droplet radius can be obtained

\begin{equation}
\lambda \,R^2\,\frac{dR}{dt} = \frac{8c_0DV}{\pi} \int_0^\infty e^{-Du^2t}\frac{1}{\left[ J_0^2(Ru) + N_0^2(Ru) \right]}\frac{du}{u},
\end{equation}
which gives a general solution for $R$ as a function of the time.
We are interested in the short and long time solutions for the concentration, the total flux of monomers at the perimeter of the droplet and the growth rate of the droplet radius.

For small values of the time, it is shown that the behaviours of $c(r,t)$ and $ \partial c(r,t)/\partial r$ may be determined from the behaviors of $\bar c(r,s)$ and $\partial \bar c(r,s)/\partial r$, respectively, for large values of the transformed parameter $s$. Then, we expand the Bessel functions occuring in (\ref{cbarad},\ref{dcbarad}) supposing s to be large. The final result for the concentration of monomers, keeping the leading time dependent term, is

\begin{equation}\label{cstad}
c(r,t) \simeq c_0\left[ 1 - \left(\frac{R}{r} \right)^{1/2} Erfc\left(\frac{r-R}{2\sqrt{Dt}}\right)\right].
\end{equation}
The total flux at the droplet perimeter and the growth rate of the droplet radius also in this limit using (\ref{intakedef}) and (\ref{growthdef}), respectively, are given by

\begin{equation}\label{fstad}
\Phi (t) \simeq 2\,c_0VR\sqrt {\pi D}\,t^{-1/2},
\end{equation}

\begin{equation}\label{Rstad}
R(t)\simeq \left(\frac{8\,c_0V\sqrt {\pi D}}{\lambda}\right)^{1/2}t^{1/4}.
\end{equation}
We see that in the short time limit, R grows as a power of the time with an exponent of $1/4$. 

For large values of the time, the behaviours of $c(r,t)$ and $\partial c(r,t)/\partial r$ may be determined from the bahaviours of $\bar c(r,t)$ and $\partial \bar c(r,s)/\partial r$, respectively, for small values of the transform parameter $s$. We then expand the Bessel functions occuring in (\ref{cbarad},\ref{dcbarad}) supposing s to be small. Keeping the leading time dependence term, the concentration of monomers yields

\begin{equation}\label{cltad}
c(r,t) \simeq 2\,c_0 \,\displaystyle\frac{\ln\left(\displaystyle\frac{r}{R}\right)}{\ln\left(\displaystyle\frac{4Dt}{\sigma ^2R^2}\right)},
\end{equation}
where $\sigma = e^\gamma = 1.78107...$, where $\gamma = 0.57722...$ is Euler's constant. The total flux at the droplet perimeter and the growth rate of the droplet radius also in this limit using (\ref{intakedef}) and (\ref{growthdef}), respectively are given by

\begin{equation}\label{fltad}
\Phi (t) \simeq 4\pi c_0 DV\left[\ln\left(\frac{4Dt}{\sigma ^2R^2}\right) \right]^{-1},
\end{equation} 

\begin{equation}\label{Rltad}
R(t) \simeq A\left[\frac{\tau}{\ln(\tau)}\right]^{1/3},
\end{equation}
where $A = \left( 9\pi V\sigma ^2/\lambda \right)^{1/3}$ and $\tau =  4c_0Dt/\sigma ^2 $ is the dimensionless time. Up to a constant, these are the same results which were obtained by Krapvisky based on a quasistatic approximation using a similarity variable approach \cite{krapquasi}.

\section{Phenomenological Rate Equation Model}

One can consider various modification of the initial and boundary conditions (2) and (4). Here we improve the model and incorporate effects other than the irreversible adsorption at $r = R$ expressed by (4), Ref. \cite{mozy}. In this section we consider a phenomenological modification of the boundary condition (4) to allow for detachment. This was introduced in \cite{mozy} where the relation

\begin{equation}
\label{pmbc}
\frac{\partial c(r,t)}{\partial t} = -\,m\,c(r,t) + k
\end{equation}
at $r=R$ was replaced for (4). Here it is assumed that the diffusing monomers that reach the perimeter of the droplet are incorporated in the aggregate structure at the rate $mc$ proportional to their concentration at $R$. The second term in (\ref{pmbc})
corresponds to detachment and is assumed that only depends on the internal processes, so there is no dependence on the external diffuser concentration \cite{mozy}.

To solve (1) with (2,3) and (\ref{pmbc}), we go through steps similar to section 2 and only emphasize the final expressions. In the Laplace transform version, the boundary condition becomes 

\begin{equation}
 (s + m) \,\bar c(r,s) = \frac{k}{s} + c_0
\end{equation}
at $r=R$. The concentration and the radial gradient of the concentration in this version become

\begin{equation}\label{cbarpm}
\bar c(r,s) = \frac{c_0}{s} - \frac{mc_0-k}{s(s+m)} \,\frac{K_0(qr)}{K_0(qR)},
\end{equation}

\begin{equation}\label{dcbarpm}
\frac{\partial \bar c(r,s)}{\partial r} = \frac{mc_0-k}{(Ds)^{1/2}} \,\frac{1}{(s+m)} \,\frac{K_1(qr)}{K_0(qR)}.
\end{equation}
Now we look for the solutions in the short and long time limits.

For small values of the time, we use the asymptotic expansions of the Bessel functions in (\ref{cbarpm},\ref{dcbarpm}) for large values of $s$ and ignore $m$ in comparison to $s$ in the term $(s+m)$. Then, the concentration, the total flux at the droplet perimeter and the growth rate of the droplet radius in this limit, keeping only the leading time-dependent terms, are given by

\begin{equation}\label{cstpm}
c(r,t) \simeq c_0 + 4mt\left( c_0-\frac{k}{m}\right)\left( \frac{R}{r}\right)^{1/2} \, Erfc\left(\frac{r-R}{2\sqrt {Dt}}\right),
\end{equation}

\begin{equation}\label{fstpm}
\Phi (t) \simeq 4mVR\sqrt{\pi D}\left( c_0-\frac{k}{m}\right)t^{1/2},
\end{equation}

\begin{equation}\label{Rstpm}
R(t)\simeq \left[ \frac{16mV}{3\lambda}\sqrt{\pi D}\left(c_0-\frac{k}{m}\right)\right]^{1/2}t^{3/4}.
\end{equation}
We see that in a phenomenological model,
R grows as a power of the time with an exponent of $3/4$ in the short time limit.
In the expressions (\ref{cstpm}-\ref{Rstpm}), in comparison with (\ref{cstad}-\ref{Rstad}) in the previous section, there is a term as $(c_0-k/m)$ which shows a reduction of the rate due to detachment, proportional to the ratio $k/m$. 

For large values of the time, we use the expansions of the Bessel functions in (\ref{cbarpm},\ref{dcbarpm}) supposing $s$ to be small and ignore $s$ in comparison with $m$ in the term $(s+m)$. Then, the concentration, total flux at the droplet perimeter and growth rate of the droplet radius, keeping only the leading time dependent terms, yield

\begin{equation}\label{cltpm}
c(r,t)\simeq \frac{k}{m}+2 \left (c_0-\frac{k}{m}\right)\displaystyle\frac{\ln\left(\displaystyle\frac{r}{R}\right)}{\ln\left(\displaystyle\frac{4Dt}{\sigma ^2R^2}\right)},
\end{equation}

\begin{equation}\label{fltpm}
\Phi (t)\simeq 4\pi DV\left(c_0-\frac{k}{m}\right)\left[\ln\left(\frac{4Dt}{\sigma ^2R^2}\right)\right]^{-1},
\end{equation}

\begin{equation}\label{Rltpm}
R(t)\simeq A\left[\frac{\tau}{\ln(\tau)}\right]^{1/3},
\end{equation}
where $A=(9\pi V\sigma ^2/\lambda)^{1/3}$ and $\tau = 4Dt(c_0-k/m)/\sigma ^2$. These asymptotic expressions are quite similar to the long time forms (\ref{cltad}-\ref{Rltad}) in section 2. The only change is the reduction of the rate due to the detachment, proportional to the ratio $k/m$.

For a fast enough detachment, the total fluxes of the monomers at the boundary in the both short and long time limits (\ref{fstpm},\ref{fltpm}) can actually become negative. In this case, the flux does not feed the growth of the droplet and the droplet volume does not increase anymore. Therefore, the mass conservation (\ref{growthdef}) does not hold and the growth laws (\ref{Rstpm},\ref{Rltpm}) are not valid anymore.  
For a case in which $c_0=k/m$, the system reaches a stationary state and therefore the total rate and the total flux of the monomers at the droplet perimeter, become zero for all times. Consequently, there is no growth for the droplet and the concentration of the monomers is equal to the initial concentration, $c_0$, for all times. These results can be obtained from the both short and long time expressions (\ref{cstpm}-\ref{Rstpm}) and (\ref{cltpm}-\ref{Rltpm}), respectively.

\section{Radiation Boundary Condition}

In this section we consider another modification of the boundary condition (4) and replace it with a radiation boundary condition

\begin{equation}
\alpha \,\frac{\partial c(r,t)}{\partial r}+\beta = c(r,t)
\end{equation}
at $r=R$, Ref. \cite{mozy}. Here it is assumed that the concentration is proportional to its derivative, with an additional constant $\beta$. Again we go through steps similar to the section 2 and only emphasize the final expressions. In the Laplace transform version, the boundary condition becomes

\begin{equation}
\alpha \,\frac{\partial \bar c(r,s)}{\partial r}+\frac{\beta}{s} = \bar c(r,s)
\end{equation}
at $r=R$. The concentration and its radial gradient in this version become

\begin{equation}\label{cbarra}
\bar c(r,s)=\frac{c_0}{s}-\frac{(c_0-\beta)}{s} \,\frac{K_0(qr)}{K_0(qR)+\alpha qK_1(qR)},
\end{equation}

\begin{equation}\label{dcbarra}
\frac{\partial \bar c(r,s)}{\partial r}=\frac{(c_0-\beta)}{(Ds)^{1/2}} \,\frac{K_1(qr)}{K_0(qR)+\alpha qK_1(qR)}.
\end{equation}
We concentrate our attention to the solutions in the short and long time limits.

For small values of the time, we use the asymptotic expansions of the Bessel functions in (\ref{cbarra},\ref{dcbarra}) to get the leading time dependent terms for the concentration, the total flux and the droplet growth rate

\begin{equation}\label{cstra}
c(r,t)\simeq c_0-\frac{2i}{\alpha}(c_0-\beta)\left(\frac{DR\,t}{r}\right)^{1/2}Erfc\left(\frac{r-R}{2\sqrt {Dt}}\right),
\end{equation}

\begin{equation}\label{fstra}
\Phi (t)\simeq \frac{2\pi RDV}{\alpha}(c_0-\beta),
\end{equation}

\begin{equation}\label{Rstra}
R(t)\simeq \left[\frac{4\pi DV}{\alpha \lambda}(c_0-\beta)\right]^{1/2}t^{1/2}.
\end{equation}
The term $(c_0-\beta )$ in these expressions shows a reduction of the rate due to the detachment, proportional to the ratio $\beta $.
We see that in the short time limit, the total flux at the droplet perimeter is time-independent and $R$ grows as a power law with an exponent equal to $1/2$.

For large values of the time, we expand the Bessel functions in (\ref{cbarra},\ref{dcbarra}) supposing $s$ to be small. Consequently, the asymptotic expressions for the concentration, total flux and the droplet growth rate, yield

\begin{equation}\label{cltra}
c(r,t)\simeq \beta + 2(c_0-\beta)\displaystyle\frac{\ln\left(\displaystyle\frac{r}{R}\right)}{\ln\left(\displaystyle\frac{4Dt}{\sigma ^2R^2}\right)},
\end{equation}

\begin{equation}\label{fltra}
\Phi (t)\simeq 4\pi DV(c_0-\beta)\left[\ln\left(\frac{4Dt}{\sigma ^2R^2}\right)\right]^{-1},
\end{equation}

\begin{equation}\label{Rltra}
R(t)\simeq A\left[\frac{\tau}{\ln(\tau)}\right]^{1/3},
\end{equation}
where $A=(9\pi V\sigma ^2/\lambda)^{1/3}$ and $\tau = 4Dt(c_0-\beta )/\sigma ^2$.
These long time expressions have the same forms as (\ref{cltpm}-\ref{Rltpm}) provided we identify

\begin{equation}
\beta = \frac{k}{m}.
\end{equation}
For a fast enough detachment, analogue to the section 3, the total fluxes of the monomers at the boundary (\ref{fstra},\ref{fltra}) can become negative. In this case, the growth laws (\ref{Rstra},\ref{Rltra}) do not hold anymore. 
For a case in which $c_0=\beta $, analogue to the section 3, the system reaches a stationary state and therefore the total flux and the droplet growth rate become zero. The concentration also does not change with the time and is equal to the initial one. These can be seen from the both short and long time results (\ref{cstra}-\ref{Rstra}) and (\ref{cltra}-\ref{Rltra}), respectively.

\section{Constant Flux Boundary Condition}

In this section we impose a condition on the flux of the monomers assuming that the total flux of monomers at the droplet perimeter is constant. Therefore, we replace (4) with

\begin{equation}\label{cfbc}
\Phi (t)=Q
\end{equation}
at $r=R$, where $\Phi (t)$ is given by (\ref{intakedef}) and $Q$ is a constant. The analogue to the previous sections, in the Laplace transform version, the boundary condition becomes

\begin{equation}
2\pi RDV\,\frac{\partial \bar c(r,s)}{\partial r}=\frac{Q}{s}
\end{equation}
at $r=R$. The concentration and its radial gradient in this version are

\begin{equation}\label{cbarcf}
\bar c(r,s)=\frac{c_0}{s}-\frac{Q}{2\pi RVD^{1/2}} \,\frac{K_0(qr)}{s^{3/2}K_1(qR)}
\end{equation}
and

\begin{equation}\label{dcbarcf}
\frac{\partial \bar c(r,s)}{\partial r}=\frac{Q}{2\pi  RDV} \,\frac{K_1(qr)}{sK_1(qR)}.
\end{equation}
Appropriate expansions of the Bessel functions in (\ref{cbarcf}) give us the limiting forms of the concentrations. For small values of the time it yields

\begin{equation}
c(r,t)\simeq c_0-\frac{i\,Q}{\pi V}\left(\frac{t}{DR\,r}\right)^{1/2}Erfc\left(\frac{r-R}{2\sqrt {Dt}}\right)
\end{equation}
and for large values of the time it gives

\begin{equation}
c(r,t)\simeq c_0-\frac{Q}{4\pi DV}\,\ln\left(\frac{4Dt}{\sigma r^2}\right).
\end{equation}
The trivial solution for the droplet growth rate using
(\ref{growthdef},\ref{cfbc}) is

\begin{equation}
R(t)=\left(\frac{3\,Q}{\lambda}\right)^{1/3}t^{1/3},
\end{equation}
for all times.

\section{Conclusions}

We studied the growth of a single, motionless, three-dimensional droplet that accommodates monomers at its perimeter on a 2D substrate. The noncoalescing monomers diffuse and are adsorbed at the aggregate perimeter of the droplet with different boundary conditions. Models with adsorption and radiation boundary conditions, and a phenomenological model for the boundary condition, were considered and solved in a quasistatic approximation. In a model with adsorption boundary condition, the droplet forms an absorber and the concentration of the monomers at its perimeter is zero. In a phenomenological model, we assumed that the diffusing monomers that reach the perimeter of the droplet, are incorporated in the aggregate structure at a rate proportional to their concentration at the boundary. We also added another term which corresponds to detachment. In a model with radiation boundary condition we assumed that the concentration is proportional to its derivative with an extra detachment term.
For each model, we solved exactly the diffusion equation for the concentration of the monomers, subject to a {\it fixed} boundary. Then, using a mass conservation law at the perimeter of the droplet, we found an expression for the growth rate of the {\it moving} boundary. Models were subjected to an initial, uniform concentration of monomers.
Asymptotic results for the concentration, total flux of monomers  at the boundary and the growth rate of the droplet radius, were obtained in both short and long time limits.
The results revealed that in both phenomenological and radiation models, in comparison with adsorption model, there is a reduction of the rate due to the detachment. The rate can become negative if the detachment is fast enough. In this case, the total flux of the monomers at the perimeter of the droplet become negative. Therefore, the flux does not feed the growth of the droplet volume and the droplet growth laws obtained in the models, are not valid anymore. For a value of the detachment for which the total rate and therefore the total flux become zero, the system reaches a stationary state and there is no growth for the droplet anymore.
The same reduction of the rate was obtained in \cite{mozy} where incorporation of particle detachment in Smoluchowski model of colloidal growth, was considered.

The results in the short time limit predicted that the radius of the droplet grows as a power of the time with different exponents for different boundary conditions. The exponents of the power laws were $1/4$, $1/2$ and $3/4$, respectively, for the models with adsorption, radiation and phenomenological boundary conditions. We see that the growth rate is the slowest for the adsorption boundary condition and is the fastest for the phenomenological model. This is because, as was said before, in the phenomenological model, the diffusing monomers at the perimeter of the droplet are incorporated in the aggregate structure of the droplet. The total flux of the monomers at the droplet perimeter is also power law with exponents of $-1/2$ and $1/2$ for the adsorption and phenomenological model, respectively, and is a constant for the radiation model. Again the flux is maximum for the phenomenological model and is minimum for the adsorption model.    

In the long time limit, the growth law for the radius of the droplet was the same for all boundary conditions. Also the concentration and total flux had the same time dependency in all models. The only change, as we said before, was the reduction of the rate due to the detachment in the both phenomenological and radiation models in comparison with adsorption model. Asymptotic results for large values of the time exhibited that the radius of the droplet increases as $[t/\ln(t)]^{1/3}$ in all models. This was obtained by Krapivsky \cite{krapquasi} where a similarity variable approach was used to treat the growth of a droplet with an adsorption boundary condition based on a quasistatic approximation.   

We saw that the time dependency of the results was the same for all the models in the long time limit and was different for different models in the short time limit. This suggests that initially the flux of the monomers at the boundary and therefore the droplet growth rate, are affected by the condition at the boundary. But in the long time limit, the system reaches a stable state and the initial effects can be ignored, therefore all the models give the same results. This suggests that a rate as $[t/\ln(t)]^{1/3}$ is a universal asymptotic growth law for the radius of the droplet independent of the boundary conditions.   

In the both models with phenomenological and radiation boundary conditions, similar to the results in \cite{mozy}, the value of the concentration at $r=R$ for large times, see (\ref{cltpm}) and (\ref{cltra}), is exactly equal to $\beta =k/m$, independent of $R$. This suggests that as far as large $R$ and large time behaviours are concerned, we can use the boundary condition

\begin{equation} 
c=\beta =\frac{k}{m}
\end{equation}
at $r=R$, instead of phenomenological and radiation boundary conditions. Indeed, the value of the concentration at $r=R$ is the only parameter needed to calculate the modifications of the asymptotic behaviours due to the detachment. With this boundary condition for all times, the asymptotic results (\ref{cltpm}-\ref{Rltpm}) and (\ref{cltra}-\ref{Rltra}) become exact. Therefore, a constant concentration of the monomers at $r=R$ for all times, gives an exact growth rate as $[t/\ln(t)]^{1/3}$ .

We also examined another model with a constant flux of monomers at $r=R$. The results showed that the radius of the droplet grows as $t^{1/3}$ for all times. Thus, the growth laws predicted by a constant concentration and by a constant flux at the boundary, differ from each other by a slowly varying logarithmic factor.

\end{document}